\documentclass[two column,showpacs,preprintnumbers,amsmath,amssymb]{revtex4}

\usepackage{graphicx}
\usepackage{dcolumn}
\usepackage{bm}

\begin{document}


\title{In an Attempt to Introduce Long-range Interactions into
\\ Small-world Networks}

\author{Chun-Yang Wang}
\thanks{Corresponding author. Electronic mail: wchy@mail.bnu.edu.cn}
\affiliation{Department of Physics and engineering, Qufu Normal
University, Qufu, 273165, China}

\date{\today}

\begin{abstract}
Distinguishing the long-range bonds with the regular ones, the
critical temperature of the spin-lattice Guassian model built on two
typical Small-world Networks (SWNs) is studied. The results show
much difference from the classical case, and thus may induce some
more accurate discussion on the critical properties of the
spin-lattice systems combined with the SWNs.
\end{abstract}

 \maketitle

\section{\label{sec:level1}INTRODUCTION}
Since the precursive study of D.J.Watts and H.E. Strogatz on
Small-world Network (SWN)\cite{s1}, there have been enormous
research activities concerning the critical properties of some
spin-lattice models built on them[2-13]. Among the various model
systems, the Guassian model\cite{s14} is always chosen as a
special one to give detailed investigation on some important
problems because of its relative simplicity in mathematics. In the
previous studies, due to the complexity of the problem itself and
the often formidable mathematical task, two spins connected by a
long-range bond are usually supposed to act in the same way as
those connected by a regular one. In other words, their
interaction contributes to the system hamiltonian to the same
extent and they obey the redistribution mechanism. In short, there
is no difference between a long-range bond and a regular one.
However, one may be curious about the extent to which the features
of phase transition will be different if one chooses certain
physical quantities such as $J$ in the Guassian model to be
different. Also prompted by the recent studies of the critical
properties of the system under long-range
interactions\cite{s15,s16}, it seems quite reasonable and
meaningful to make such an attempt on the SWN, which is always
looked as the model systems for the networks in reality.

The present study is thus aimed to make an investigation on the new
features of the critical temperature of the Guassian model built on
SWNs caused by the distance dependent power-law decaying long-range
interaction $(J\propto r^{-\alpha})$. It is reported in such a
stepwise sequence: Sec.II presents our definition of the
spin-lattice Guassian model on SWNs and the certification of its
feasibility to be categorized into the two main groups of SWNs being
studied. The succeeding one embodies the thematic study of how the
critical temperature is affected by the long-range interactions.
Sec.IV gives the summarization with some discussions.

\section{SPIN-LATTICE GUASSIAN MODEL BUILT ON SMALL-WORLD
NETWORKS}

Following the first prototype of Small-world Network and a number
of other variants\cite{s1,s11,s17,s18}. we define the spin-lattice
Guassian model built on SWNs in such a little different way:
starting with an initial $D-$dimensional network of $N$ spins with
periodic boundary condition, each spin is linked to its $2kD$
 nearest neighbors (we choose $k=1$ in this paper). Then (1)
 without changing the original bonds, but with
 probability $p_{1}$, each pair of non-nearest neighboring spins is
 additionally connected by a ``shortcut''(As a model system for
 networks in reality, we suppose the number of shortcut to be
 much smaller than $N$. Practically we require $Np_{1}\ll 1$);
  (2) the spins are visited one after another, and each bond
   connecting a spin to one of its $k$ nearest neighbors is reconnected
   with probability $p_{2}$ to another randomly chosen one. In the sequel,
    by comparing some of the results we got with those got in other studies,
    we will show how the modelling systems defined above can be
    categorized into the typical two main groups of SWNs. Let us
    begin with the classical Guassian effective Hamiltonian

\begin {equation}
H=\frac{1}{2}\sum_{mn}K(r_{mn})\sigma_{m}\sigma_{n}-\frac{b}{2}\sum_{m}\sigma_{m}^{2},
\end {equation}
where $K(r_{mn})={J(r_{mn})}/{k_{B}T}$ is the reduced distance
dependent interaction between a spin pair. The spins can take any
real value between $(-\infty,+\infty)$, and the probability of
finding a given spin between $\sigma_{m}$ and $\sigma_{m} + d
\sigma_{m}$ is assumed to be the Guassian-type distribution
$p(\sigma_{m})d\sigma_{m}\propto[\exp-(b/2) \sigma_{m}^{2}]
d\sigma_{m}$. $b$ is the Guassian distribution constant. $k_{B}$
the Boltzman constant and $T$ the thermodynamic temperature.

Generally the critical point of a Guassian system is determined by
the singularity of the free energy

\begin{equation}
F=\frac{1}{2}k_{B}T\lim_{N\rightarrow\infty}N^{-1}\sum_{\vec{q}}\ln(b-K(\vec{q}))+T\cdot
C ,
\end{equation}
where $C$ is a constant. However, since $K(\vec{q})$ here must not
take any value greater than $b$, one can always obtain the
critical point by

\begin {equation}
K_{max}(\vec{q})=b.
\end {equation}
here $K_{max}(\vec{q})$ is the Fourier transform of $K(r_{mn})$

\begin {equation}
K(\vec{q})=\sum_{\vec{r}_{mn}}K(r_{mn})e^{i\vec{q}\cdot\vec{r}_{mn}}.
\end{equation}
Supposing firstly there is no difference between along-range bond
and a regular one, that is, $K(\vec{q})$ is a constant $K$
independent of $r_{mn}$, Eq.(4) then becomes

\begin{equation}
K(\vec{q})=K\sum_{\vec{r}_{mn}}e^{i\vec{q}\cdot\vec{r}_{mn}}.
\end{equation}

For the one-dimensional Guassian chain constructed by the
shortcut-adding operation introduced at the beginning of this
section, one can easily find $\vec{r}_{mn}=\pm Ma\vec{i}$. where
$M$ is the minimum number of the intervals between any two spins.
See Fig.1 (a) for an example, the value of corresponding to the
two nodes $A$ and $B$ is defined as 3, and that between nodes $B$
and $C$ is 2. Thus $M$ can take any integral value between
$[1,(N-1)/2]$. where $M=1$ correspond to the case of no
``shortcut'' has been added into the network. $a$ is the lattice
constant.

\begin{figure}
\includegraphics[scale=0.7]{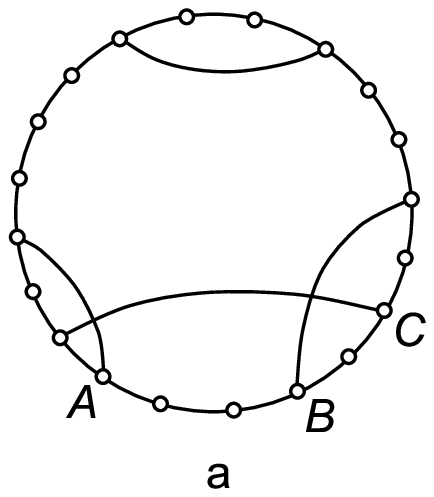}
\includegraphics[scale=0.7]{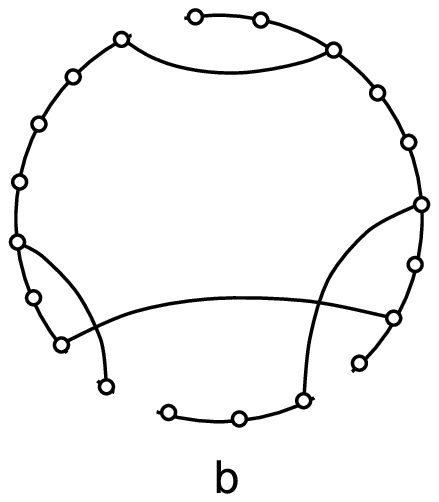}\caption{Map of the exhibition
 for an adding-type(a) or rewiring-type (b) SWN at $N=20$}
\end{figure}

Now Eq.(5) can be expressed as

\begin{eqnarray}
K(\vec{q})=2K\cos(qa)+2Kp_{1}\cos(2qa)+ \nonumber \\\cdots
+2Kp_{1}\cos((N-1)qa/2).
\end{eqnarray}
setting $q=0$, $K(\vec{q})$ then reaches its maximum
\begin{equation}
K_{max}(\vec{q})=K(2+(N-3)p_{1}).
\end{equation}
By similar analysis this result can be extended to systems with
higher dimensionality as
\begin{equation}
K_{max}(\vec{q})=K(2d+(N-3)p_{1}),
\end{equation}
where $d$ is the spatial dimension. Then the general expression of
the critical point is obtained to be
\begin{equation}
K_{C}=\frac{b}{2d+(N-3)p_{1}}.
\end{equation}
Where $K_{C}$ is the value of $K$ when the system reach its
critical point.

In order to determine the critical point of the Guassian system
constructed by the second kind of operation, we should take into
account two important factors. Firstly, the bond-rewiring
operation haven't altered the total number of bonds in the network
but it will change the coordination number of some certain nodes.
Secondly, the rewiring operation makes the structure of the
network rather irregular. For a convenience aim, we suppose the
bonds rewired are reconnected to the modes distributing
symmetrically around a reference one $m$. At the same time, notice
that the coordination number of the lattice is still 2 on average
although the neighbors of some nodes can be greater or smaller
than it. In this mean-field view, we get

\begin{equation}
K(\vec{q})=2K(1-p_{2})\cos(qa)+2Kp_{2}\cos(qMa).
\end{equation}
where $M$ ranges from 2 to $(N-1)/2$. Thus we obtain the critical
point: $K_{C}=b/2$, and in case of higher dimensionality

\begin{equation}
K_{C}=\frac{b}{2d}.
\end{equation}

Seen from Eqs.(9) and (11), the critical point of the Guassian
sytem built on the two kinds of networks we defined inclines to
the same values as those got in earlier studies\cite{s11}.
Therefore, we can undoubtedly categorize the networks we defined
to the two main groups of SWNs-the adding-type SWN
\cite{s11,s17,s18}and the rewiring-type SWN\cite{s1,s11}
respectively.

\section{CRITICAL TEMPERATURE OF THE SYSTEM UNDER LONG-RANGE INTERACTIONS}

The distance dependent power-law decaying long-range interaction
ubiquitously exists in reality and is relatively easy to treat, so
with an aim to give the study rather qualitatively than
quantificationally, the choose of such a form of interaction is
naturally. From the point of view of theoretical physics, this
means to set

\begin{equation}
J(r_{mn})=J\cdot(\frac{r_{mn}}{a})^{-\alpha},
\end{equation}
where $\alpha >0$ is the decaying rate constant.

Now let's go on to see how and to what extent the long-range
interactions affect the critical temperature of the system.
Firstly, we will aim at the one-dimensional Guassian model built
on the adding-type SWN. In accordance with Eq.(12), $K(r_{mn})$
can take two separate forms of values. One is $K(r_{mn})=K$, which
represents the interactions originally exist between the
nearest-neighboring spin pair. The other is
$K(r_{mn})=K/M^{\alpha}$ representing the interaction between two
spins connected by a long-range bond, where $M\in[2,(N-1)/2]$, and
the distance between two spins is supposed to be proportional to
the lattice constant $a$. Not to forget here that, the long-range
bond was added in with the probability of $p_{1}$. So after
Fourier transform $K(r_{mn})$ we get
\begin{equation}
K(\vec{q})=2K\cos(qa)+\sum_{M=2}^{(N-1)/2}\frac{2K}{M^{\alpha}}p_{1}\cos(qMa),
\end{equation}
and its maximum
\begin{equation}
K_{max}(\vec{q})=2K+\sum_{M=2}^{(N-1)/2}\frac{2K}{M^{\alpha}}p_{1}.
\end{equation}

Then the criticalpoint of the system can be determined easily
through Eq.(3). For analytical convenience, we show in Fig.(2) the
$\alpha$ and $p_{1}$ dependence of $K_{C}$ for a certain network
with $N$ spins. Seen from it $K_{C}$ increases sharply with
$\alpha$, approaching quickly to the stable value $K_{C}=b/2$,
which corresponds to the result of only considering
nearest-neighboring interactions. This means the affection of the
long-range interaction on the critical temperature is most
apparent at very small values of $\alpha$. If $\alpha$ take a
large value, the long-range interaction will quickly attenuate and
show hardly any influence on the critical temperature.

In the succession, let's treat the one-dimensional Guassian model
on the rewiring-type SWN. By a similar analysis, one can easily
find $K(r_{mn})$ also takes two forms of values here. They are:
$K(r_{mn})=K$ with the probability of $1-p_{2}$; and
$K(r_{mn})=K/M^{\alpha}$ with the probability of $p_{2}$. And the
value of $M$ still range from 2 to $(N-1)/2$. Thus we can express
$K(\vec{q})$ as

\begin{equation}
K_{max}(\vec{q})=2K(1-p_{2})\cos(qa)+\sum_{M=2}^{(N-1)/2}\frac{2K}{M^{\alpha}}p_{2}\cos(qMa).
\end{equation}
and its maximum

\begin{equation}
K_{max}(\vec{q})=2K(1-p_{2})+\sum_{M=2}^{(N-1)/2}\frac{2K}{M^{\alpha}}p_{2}.
\end{equation}

The $\alpha$ and $p_{2}$ dependence of $K_{C}$ can be got from
Eqs.(3) and (16) without difficulty, shown in Fig.(3). Notice that
there is an interesting feature of the curves: They form an
intersection at $\alpha\cong1.75$, which means $K_{C}=b/2$ for
arbitrary values of $p_{2}$. In a physical view, the bond-rewiring
operation have no affection on the critical temperature. This is
in accordance with the results reported in earlier studies. But it
does not always make sense. If we take $\alpha>1.75 (<1.75)$ the
critical temperature decreases (increases) respectively, and when
$\alpha\rightarrow\infty$ it tends to be a $p_{2}$ dependent
constant. This shows that practically the critical temperature of
the Guassian system on the rewiring-type SWNs will be altered by
the long-range interactions. It may be a result of the totally
different role played by the long-range bonds.

\begin{figure}
\includegraphics[scale=0.3]{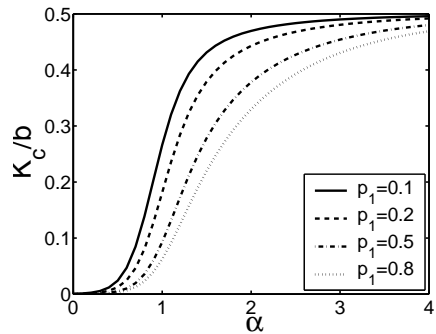} \caption{The $\alpha$ and $p_{1}$ dependence of $K_C/b$ on the
 adding-type SWNs, taking
 into account the affect of long rang interactions. Where $N=2000$.}
\end{figure}

\begin{figure}
\includegraphics[scale=0.3]{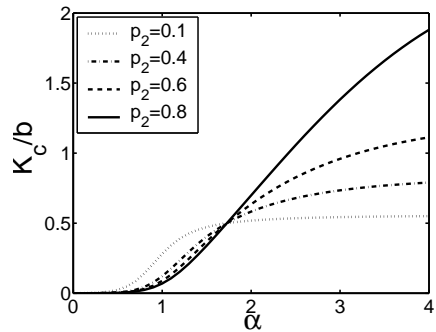} \caption{The $\alpha$ and $p_{2}$ dependence of $K_C/b$ on the
 rewiring-type SWNs, taking
  into account the affect of long rang interactions. Where $N=2000$.}
\end{figure}

\section{SUMMARY AND DISCUSSION}

 In this paper, we made an attempt to study the critical
 temperature of the spin-lattice Guassian model built on the two
 groups of typical Small-world Networks affected by the power-law
 decaying long-range interaction. The results we got qualitatively
 show the new features of the critical temperature caused by the
 long-range interactions: On the adding-type SWN, the present of
 the long-range bonds increased the contact of a spin with the
 system, assisted the system to behave as a whole, and thus
 increased the critical temperature of the system to some extent. While on the
 rewiring-type SWN, the long-range interactions partly replaced
 the nearest-neighbor coupling, the change of the critical
 temperature then mostly depends on the competition of such two
 kinds of interactions. Anyone's superiority will lead to the
 increase or decrease of the critical temperature. If they
 contribute to the system's Hamiltonian equally in extent, the
 critical point will remain as the classical result
 $K_{C}=b/2$.

 We have to admit that the present study is not strictly accurate
 in calculating the critical temperature, but it indeed satisfied
 the intention to give a qualitatively analysis on it. We hope that
 further studies on the critical properties of spin-lattice models
 built on SWNs will continue to reveal interesting topics of the
 widely existing critical phenomena combined with the Small-world
 Networks.

\section * {ACKNOWLEDGEMENT}
This work was supported by the Scientific Research Starting
Foundation of Qufu Normal University and the National Natural
Science Foundation of China under Grant No. 10847101.

\end{document}